\documentclass{aastex63}

\usepackage{amsmath}

\received{}
\revised{}
\accepted{June 2 2022}
\submitjournal{ApJL}

\shorttitle{A megaton-class impact flash on Jupiter}
\shortauthors{Arimatsu et al.}

\begin{document}

\title{Detection of an extremely large impact flash on Jupiter by high-cadence multiwavelength observations}

\correspondingauthor{Ko Arimatsu}
\email{arimatsu.ko.6x@kyoto-u.ac.jp}

\author[0000-0003-1260-9502]{Ko Arimatsu}
\affiliation{The Hakubi center / Astronomical Observatory, Graduate School of Science,  Kyoto University \\
Kitashirakawa-oiwake-cho, Sakyo-ku, Kyoto 606-8502, Japan}

\author[0000-0001-7143-6520]{Kohji Tsumura}
\affiliation{Department of Natural Science, Faculty of Science and Engineering, Tokyo City University, Setagaya, Tokyo 158-8557, Japan}

\author[0000-0003-2273-0103]{Fumihiko Usui}
\affiliation{Department of Space Astronomy and Astrophysics, Institute of Space and Astronautical Science (ISAS), Japan Aerospace Exploration Agency (JAXA),  3-1-1 Yoshinodai, Chuo-ku, Sagamihara, Kanagawa 252-5210, Japan}

\author[0000-0003-4391-4446]{Jun-ichi Watanabe}
\affiliation{Astronomy Data Center, National Astronomical Observatory of Japan
2-21-1 Osawa, Mitaka, Tokyo 181-8588, Japan}

\begin{abstract}
We report the detection of an optical impact flash on Jupiter on 15 October 2021 by a dedicated telescope, Planetary ObservatioN
Camera for Optical Transient Surveys (PONCOTS), for the first time. Our temporally resolved three-band observations of the flash allowed investigations of its optical energy 
without the need for approximations on the impact brightness temperature.
The kinetic energy of the impactor was equivalent to approximately two megatons of TNT, an order of magnitude greater than that of previously detected flashes on Jupiter and comparable with the Tunguska impact on Earth in 1908. 
This detection indicates that Tunguska-like impact events on Jupiter occur approximately once per year, two–three orders of magnitude more frequent than terrestrial impacts. 
The observed flash displayed a single-temperature blackbody spectrum with an effective temperature of approximately 8300 K without clear temporal variation, possibly representing common radiative features of terrestrial Tunguska-class superbolides.
\end{abstract}

\keywords{Impact phenomena(779) --- Time domain astronomy(2109) --- Solar system astronomy(1529) --- Planetary atmospheres(1244) --- Jupiter(873)}

\section{Introduction}
\label{sec1}
Optical flashes caused by impacts of decameter-sized outer solar system objects on Jupiter have been observed by amateur astronomers. 
Since the first detection in 2010 \citep{Hueso2010-xv}, six impact flashes had been serendipitously observed on the surface of Jupiter by 2020 \citep{Hueso2013-ki,Hueso2018-nr,Sankar2020-lm}. Investigation of the energy and frequency of these impacts provides a unique opportunity to explore the abundance of small objects in the outer solar system, as it is impossible to detect them directly \citep{Hueso2013-ki,Hueso2018-nr,Giles2021-sj}. 
Their emission characteristics should also demonstrate the radiative consequences of decameter-sized impacts on planetary atmospheres, which could potentially threaten human society \citep{Jenniskens2019-tg,Boslough1997-pi,Boslough2008-ae} but are unknown due to their infrequent occurrence on Earth \citep{Brown2002-aq}.   

Jupiter-family comets are thought to be a primary source of Jovian impactors, with an dynamical model by \citet{Levison2000-ze} indicating that their impact rate is much higher than those of other possible asteroid populations \citep{Zahnle1998-qi,Di_Sisto2005-fs,Di_Sisto2019-iq}. 
Thus, the frequent detection of flashes indicates abundant decameter-sized cometary bodies, contrary to previous studies of surface cratering rates of Jovian satellites \citep{Zahnle2003-ew,Schenk2004-yg}. 
However, these impacts were detected only serendipitously by camera sensors using Bayer RGB or monochrome filter systems attached to a single bandpass filter \citep{Hueso2013-ki,Hueso2018-nr}. 
In the former case, 
the Bayer filter array reduces the sensitivity at the individual wavelength bands due to limited photon-collecting areas.
Detailed flash analyses thus involved data from the latter \citep{Hueso2010-xv,Sankar2020-lm}, although these provide limited spectral information due to the single-band observations. Previous studies have therefore assumed a blackbody radiation spectrum with a wide temperature range in deriving the total flash optical energy \citep{Hueso2013-ki,Hueso2018-nr,Sankar2020-lm}. Multiwavelength observations with monochrome sensors are needed to estimate the optical energies of flashes without poor approximations.

This {\it Letter} presents our discovery of an Jovian impact flash on 15 October 2021 by a observation system dedicated to flash survey. In Section~\ref{sec2}, we introduce our observation system and detailed of the detection. We present our observation results of the October 2021 impact in Section~\ref{sec3} and discussion in Section~\ref{sec4}.

\section{Observation and Data Reduction}
\label{sec2}
We observed an impact flash on Jupiter at 13:24:13 UTC on 15 October 2021 with the Planetary ObservatioN Camera for Optical Transient Surveys (PONCOTS) observation system. The PONCOTS program is dedicated to monitoring flashes on Jupiter as part of the Organized Autotelescopes for Serendipitous Event Survey (OASES) project \citep{Arimatsu2017-lg}, which aims to investigate short-timescale transients in the solar system \citep{Arimatsu2019-zk}. The PONCOTS system comprises a 0.279 m aperture Schmidt-Cassegrain telescope (Celestron C11) equipped with high-cadence monochrome complementary metal-oxide semiconductor (CMOS) cameras (QHY5III-290M camera with a SONY IMX 290 sensor for camera modules of the two shorter wavelength beams and Planetary one Neptune-CII camera with a SONY IMX464 sensor for the longest wavelength beam).
With the current setup, PONCOTS monitors Jupiter using two channels, the V ($505 \-- 650$ nm) and ${\rm CH_4}$ ($880 \-- 900$ nm) bands, simultaneously with frame rates of 40 and 10 fps, respectively. 
The central wavelength of the ${\rm CH_4}$ band ($\simeq 890$ nm) corresponds to a strong absorption band of ${\rm CH_4}$ 
where the Jovian clouds are dark \citep{Karkoschka2010-vn}, and only elevated clouds and hazes appear bright.
We note that the spectral coverage of the ${\rm CH_4}$ band is slightly broader than the bandwidth of the strong methane absorption.
Images were stored in a 16-bit SER image format together with header information of the acquisition time of each frame.
Timing between timestamps of the two cameras was well synchronized with an accuracy of less than the PONCOTS V-band single-frame exposure time (0.025 s). 
However, as the PONCOTS system does not use the Global Positional System (GPS), uncertainty in the absolute timing of the PONCOTS data can be up to a few seconds.

\begin{figure}[!pt]
\begin{center}
  \includegraphics[scale=0.85]{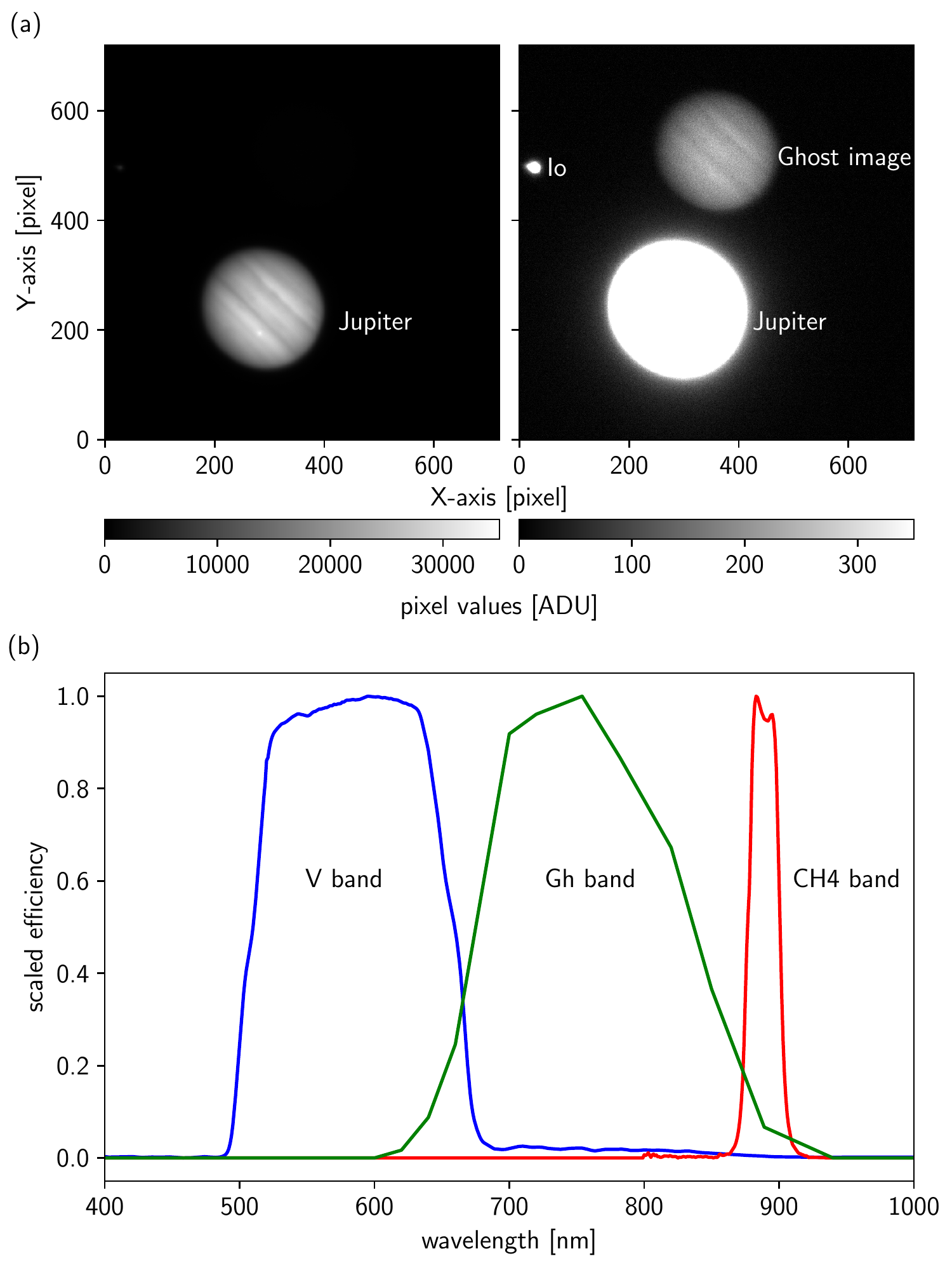}
  \caption{
  PONCOTS V-band image and spectral-system response for the three PONCOTS bands.
  (a) The left panel is an example of the PONCOTS V-band image of Jupiter obtained during the flash; right panel is as for the left but shown with a different surface brightness range. 
  An artifact “ghost” image appears in the field of view. 
  The bright spot at left in the image corresponds to Io, a satellite of Jupiter. 
  (b) Relative spectral responses of the three PONCOTS bands. 
  Blue, green, and red lines indicate the efficiency for the PONCOTS V, Gh, and ${\rm CH_4}$ bands, respectively. 
  Spectral responses for the V and ${\rm CH_4}$ bands were estimated by convolving quantum efficiencies of CMOS sensors with the filter response and transmittance or reflectance of the dichroic mirrors as provided by manufacturers. 
  The Gh band response was obtained from signal values of artifact images of the spectrophotometric standard stars at each wavelength, using narrowband filters.
  }
 \label{fig1k}
\end{center}
\end{figure}

In V-band data, an artefact of Jupiter caused by internal reflections of the dichroic mirror appears as a “ghost” image, as shown in Figure~\ref{fig1k}a. 
The intensity of artifacts generally has spectral dependence (e.g., \cite{Arimatsu2011-af}) because the degree of imperfection of the anti-reflection coating varies with wavelength.
We investigated the wavelength dependence of the artifact using datasets for spectrophotometric standard stars. 
Artifact intensity relative to that of incident light at individual wavelengths was estimated using the PONCOTS V-band camera with narrowband filters to derive the spectral dependence of artifact generation efficiency. 
We found that this image has a spectral response with an effective wavelength of 680–840 nm, different to that of the V-band image (Figure~\ref{fig1k}b). 
We thus used the ghost image as a “Gh band” image, enabling three-band simultaneous monitoring of Jupiter.

The observation system was installed on the rooftop of Building 4, 
Yoshida North Campus, Kyoto University, Kyoto, Japan. 
PONCOTS monitoring observations for the Jovian flash began on 9 September 2021. 
Until January 2022, we undertook monitoring for a total of 26.2 h.

\section{Results}
\label{sec3}

\begin{figure}[!pt]
\begin{center}
   \includegraphics[scale=0.6]{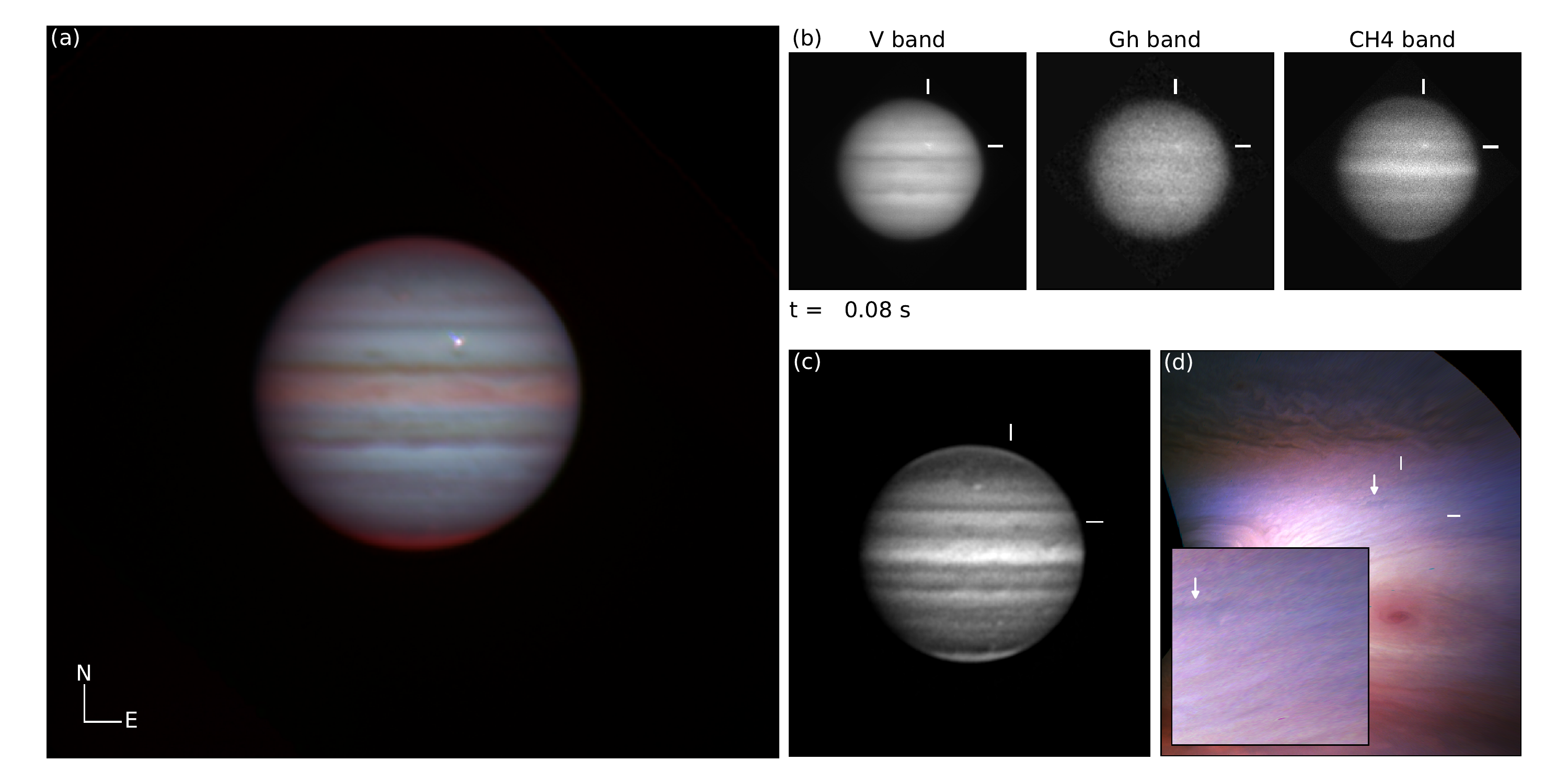}
   
   \caption{
   Impact flash on Jupiter on 15 October 2021 observed at 13:24:13 UTC. (a) Color composite image of the flash obtained with the PONCOTS three-band observations. Blue, green, and red channels of the image were constructed based on the PONCOTS V ($505-650$ nm), Gh ($680-840$ nm), and ${\rm CH_4}$ ($880-900$ nm) band images, respectively. Individual band images were processed by stacking frames obtained during the flash and adding a Jupiter image built from a stack of the frames obtained within 60 s before and after the impact. Stacked images were processed using high-pass filters to increase the contrast of surface features.
   (b) Sequential images of the flash obtained with the PONCOTS each band. Left, middle, and right panels show images obtained with the PONCOTS V, Gh, and ${\rm CH_4}$ bands, respectively. Gh band images are convolved with a Gaussian kernel to reduce noise.
   An animation of panel (b) is available. It covers within 12 seconds before and after the impact.
   (c) Image of Jupiter obtained with the PONCOTS ${\rm CH_4}$ band 16 min after impact, processed using high-pass filters to increase the contrast of surface features. 
   (d) Image of Jupiter taken by the JunoCam (red, green, and blue channels) aboard the Juno spacecraft 28 h after impact. 
   The inset shows a zoom over the impact area. 
   A slight dark structure indicated by arrows is seen close to the approximate impact position, but no noticeable impact feature is apparent.
   The pixel scale is approximately 30 km per pixel at the impact site.
   The two lines in panels (b)$\--$(d) indicate the impact location.}
 \label{fig2k}
\end{center}
\end{figure}

The three-band simultaneous observations and frequent sampling during the entire event of the 15 October 2021 flash constitute an unprecedented set of observations for an impact flash on Jupiter, as shown in Figures~\ref{fig2k}a and b. 
The impact occurred in the north tropical zone at the Jovian System III longitude and latitude of $40\degr$ and $+20\degr$, respectively. 
After the announcement of our discovery on Twitter\footnote{English version: \url{https://twitter.com/OASES_miyako/status/1449035918459879425}, Japanese version: \url{https://twitter.com/OASES_miyako/status/1449028895454367757}.}, 
we have received two reports of observations from Japanese amateur observers, “{\it yotsu}”\footnote{\url{https://twitter.com/yotsuyubi21/status/1449369747951292422?s=21&t=tAGr_6SfggpTbBqg18fGXQ}} in Aichi prefecture 
and Yasunobu Higa 
(He found the flash in an unrecorded preview screen during the Jupiter observation)
in Okinawa prefecture. 
A Singaporean amateur astronomer, Victor PS Ang, reported another observation of the flash\footnote{\url{https://www.facebook.com/groups/421163751426836/permalink/1825072731035924/}}.
These observation reports constitute simultaneous detections of the flash, which unambiguously occurred on Jupiter and not in the terrestrial atmosphere. 
The recorded movies by yotsu's and VA's observations show flash images saturated in several pixels over several frames, indicating the particularly bright nature of the flash.
Monitoring of Jupiter with PONCOTS continued for 16 minutes after the impact. 
As shown in Figure~\ref{fig2k}c, the images obtained after the impact did not show impact-debris features at the site.
Later in situ observations by the JunoCam \citep{Hansen2017-bt} aboard the Juno spacecraft were carried out approximately 28 h after impact.
Although a slight dark structure is seen close to the approximate impact site (Figure~\ref{fig2k}d), we have no evidence that this is the impact feature.

\begin{figure}[!pt]
\begin{center}
   \includegraphics[scale=0.85]{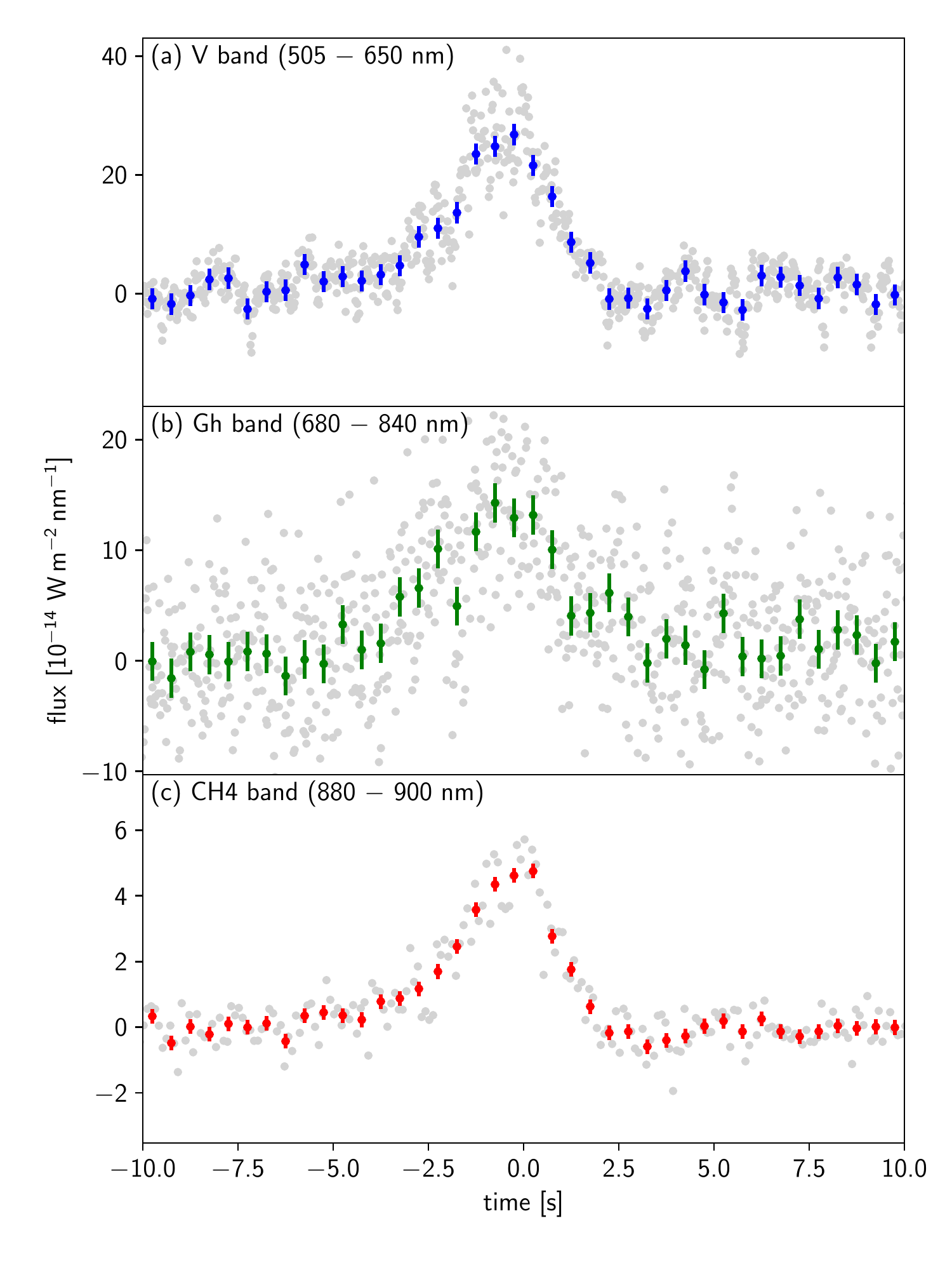}
   \caption{Light curves of the October 2021 impact flash obtained with the PONCOTS observation system. Calibrated light curves for the V, Gh, and ${\rm CH_4}$ bands are presented in panels (a), (b), and (c), respectively. Fluxes for individual frames and combinations of 0.5 s bins are shown as grey dots and points with 1$\sigma$ error bars, respectively. Background fluctuation noise is expected to dominate the error of photometry, so the 1$\sigma$ errors were estimated by examining oscillations of the light curve with no flash in the image.
  }
 \label{fig3}
\end{center}
\end{figure}

To extract the fluxes of the flash, we made differential images by subtracting stationary features from original images, constructed using pre- and post-impact images. We then performed aperture photometry on the impact location using differential images to derive the signal values of the flash. 
Signals obtained by aperture photometry of the flash were calibrated with a spectrophotometric standard star (HR 7950; V = 3.78 mag; Spectral type A1V; \cite{Hamuy1992-nw}, see Appendix~\ref{appa} for details). 
Calibrated light curves of the impact flash from PONCOTS three-band images are shown in Figure~\ref{fig3}. 
The light curves show clear features typical of previous flashes \citep{Hueso2013-ki, Hueso2018-nr},
with a slower rise phase before a peak, and a steeper decay phase. 
The apparent peak brightness of the flash was 4.7 mag in the V band, 
equivalent to an absolute magnitude 
(brightness if the object were at a distance of $10^5$ m) of –29,
approximately 300 times the brightness of the Sun as observed from Jupiter.
Furthermore, the flash was visible for $\sim 5.5$ s, longer than the previous flashes ($1 \-- 2$ s; \cite{Hueso2013-ki,Hueso2018-nr}). 
This duration is consistent with the simultaneous observation results.
A significant amount of optical energy was thus released during the impact.

As shown in Figures~\ref{fig3} and \ref{fig4}a, the observed fluxes indicate excess flux in the V and Gh bands relative to the ${\rm CH_4}$ band. The apparent spectral trend was partially caused by a wavelength-dependent reflection contribution by Jovian cloud. 
Strong backward scattering of Jovian upper clouds \citep{Heng2021-my,Li2018-uj} significantly contributed to cloud reflection in the V and Gh bands. 
To correct this, we estimated the contribution of the cloud-reflection component based on wavelength-dependent scattering phase functions of the Jovian surface provided by \citet{Heng2021-my}. 
In this estimation, we assume a flash with an altitude higher than the clouds and the methane absorption layer of the atmosphere.
The cloud-reflection component is approximately $70\%$, $60\%$, and $30\%$ of the observed fluxes in the V, Gh, and ${\rm CH_4}$ bands, respectively (Figure~\ref{fig4}a). 
Details of the procedure will be given in a separate paper.

Our first acquisition of high-cadence three-band light curves allowed investigation of time-resolved spectral characteristics of the impact flash. 
Flux data were binned into 0.5 s time bins to provide temporal spectral variations with sufficient signal-to-noise ratios (Figure~\ref{fig3}). 
Three-band spectral energy distributions (SEDs) for the time bins are shown in Figures~\ref{fig4}a and \ref{fig6} in Appendix~\ref{appb}. 
We then fitted each 0.5 s bin SED with a single-temperature blackbody-radiation spectral model. After cloud-reflection correction, SEDs for most bins were approximated by a single-temperature blackbody spectrum, at least in the peak and decay phase, where $> 70\%$ of the total optical energy was emitted. 
A slight excess in the Gh band during the rise phase implies a contribution of non-thermal components, although the statistical significance is marginal. 
The best-fit optical energy and temperature of each time bin are indicated in Figure~\ref{fig4}b, 
with the best-fit temperature being $8300 \pm 600 $ K without evident temporal variation. 
This value is consistent with the previous analyses of brightness temperatures of Jovian impact flashes ($9600 \pm 600$ K; \cite{Giles2021-sj}, $6500 \-- 8500$ K; \cite{Hueso2013-ki}, $7800 \pm 600$ K; \cite{Chapman1996-zx}).
Based on best-fit results for temporal optical energy for individual time bins, total optical energy $E_0$ was determined to be
$E_0 = 1.8^{+0.9}_{-0.2}\times 10^{15} \, {\rm J}.$
Total kinetic energy $E_T$ was derived from $E_0$ through the relationship proposed by \citet{Brown2002-aq},
\begin{equation}
E_T = 8.25 \, E_0^{0.885},
\end{equation}
where $E_T$ and $E_0$ are in kiloton TNT (kt; $1\, {\rm kt} = 4.185 \times 10^{12} \,  {\rm J}$).
We determined that $E_T={7.4}^{+3.3}_{-0.9} \times{10}^{15}\, \mathrm{J}$, approximately an order of magnitude greater than that of earlier impact-flash events ($(0.13 \-- 1.7) \times 10^{15} \, {\rm J}$; \cite{Hueso2013-ki,Hueso2018-nr,Sankar2020-lm}) 
and marginally comparable with the Tunguska explosion of 1908 ($(1 \-- 6) \times 10^{16} \, {\rm J}$; \cite{Boslough1997-pi,Boslough2008-ae}). 
The impact velocity $v_0$ for Jovian impactors is thought to be comparable with the escape velocity of Jupiter \citep{Harrington2004-kf},
$v_0 \simeq 60 {\rm km \, s^{-1}}$. 
The mass of the impactor $M_0$ was therefore estimated to be $M_0 = {4.1}^{+1.9}_{-0.5} \times 10^6 \, {\rm kg}$. 
Assuming the bulk density of the spherical impactor to be $\rho = 250$, $600$, and $2000\, {\rm kg \, m^{-3}}$, 
based on previous impact-flash studies \citep{Hueso2013-ki,Hueso2018-nr}, 
the diameters of the impactor were estimated to be $D = 31.5^{+4.2}_{-1.3}$, ${23.5}^{+3.1}_{-1.0}$, and ${15.8}^{+2.1}_{-0.6} \, {\rm m}$, respectively.

\section{Discussions}
\label{sec4}
\subsection{The lack of observed impact debris}
Hubble Space Telescope (HST) observed visible impact-debris features from the fragment N in the comet Shoemaker–Levy 9 (SL9) impact \citep{Hammel1995-nn,Crawford1997-sg}, whose mass is only $\sim$ three times larger than the estimated mass of the present impactor. 
Therefore, one would expect detection of impact features in this time as well. 
However, our follow-up observation 16 min after the impact (Figure~\ref{fig2k}c) and in-situ observation 28 h after the impact by JunoCam (Figure~\ref{fig2k}d) detected no evident debris features.

We should note that the observable timescale of the debris is unknown.
As discussed by \citet{Hueso2018-nr} based on the debris observation of SL9 fragment N, debris features associated with 50 m class impacts would be observable for several tens of hours with HST-class instruments.
Since the observable timescale of the debris features highly depends on the impactor's mass (an impactor with masses $\sim 50$ times larger would leave an observable feature with a timescale of $\geq 10^1$ times longer, \cite{Hueso2018-nr}), 
the present case might leave a very shorter-lived debris field that would become undetectable within 28 h even for the JunoCam.
We thus cannot rule out the possibility of short-timescale, 
and small-spatial-scale (undetectable by ground-based follow-ups) debris features associated with the present impact.

\subsection{Impact rate of the cometary objects on Jupiter and Earth}
Our observation is the first recorded detection of an impact flash caused by an event of total kinetic energy equivalent to $1.8^{+0.8}_{-0.2}\times{10}^3 \, {\rm kt}$, i.e., approximately two megatons of TNT (Mt). 
Estimated kinetic energies of previous impacts were $ < 400\, {\rm kt}$ \citep{Hueso2018-nr,Sankar2020-lm}. 
The event is thus the most energetic impact flash observed in the solar system since SL9 fragments impacted Jupiter in 1994 \citep{Harrington2004-kf}.

To elucidate the impact rate of Mt-class impacts on Jupiter from our single detection, we estimated the total effective observation time by all amateur surveys, following previous impact studies by \citet{Hueso2013-ki,Hueso2018-nr}.
The total effective survey time $T$ is approximated by 
\begin{equation}
T=N \, t_l \, \varepsilon \, f ,
\end{equation}
where $N$ is the number of reported observation images since 2010 archived by the Planetary Virtual Observatory and Laboratory (PVOL; \cite{Hueso2010-vi}), 
the database of solar system planets for amateur astronomers worldwide, 
$t_l$ is the duration of each observation, 
$\varepsilon$ is the efficiency of the impact detection, 
and $f$ is the conversion factor between the total observation time uploaded to PVOL 
and that of the entire amateur astronomy community.
From January 2010 to December 2021, $N$ corresponds to be $N = 31,385$ images. 
$t_l$ and $\varepsilon$ were set to 10 min and 0.4, respectively, 
which are the median values of the parameter ranges set by previous studies \citep{Hueso2013-ki,Hueso2018-nr}.
We estimated $f$ to be $25/4 = 6.25$ because four of the 25 observers of flashes were major contributors to the PVOL database. 
The total effective survey time since 2010 is estimated to be $T = 5.5 \times 10^2 \, {\rm days}$. 
Our detection thus indicates an observable occurrence rate of impacts on Jupiter with kinetic energy comparable to or greater than 1 Mt to be
$1/T = {0.7}_{-0.6}^{+1.5}$ per year.
The $1\sigma$ error range of the rate corresponds to the Poisson noise of the single detection, 
which makes the most significant contribution to the total uncertainty in the estimate.
Ground-based observations survey only the dayside of Jupiter, so the number of impacts would be two times higher, 
corresponding to $\sim 1.3_{-1.1}^{+3.1}$ per year,
which is compared with previous estimates in Figure~\ref{fig5}. 
For comparison, we also infer $T$ based on the data gathered and analyzed by an impact detection software tool, DeTeCt\footnote{\url{http://www.astrosurf.com/planetessaf/doc/project_detect.php}} \citep{Delcroix2020-cy}.
As of May 2022, DeTeCt analyzed $T = 2.4 \times 10^2$ days in total, 
and the impact rate corresponds to $\sim 3.0_{-2.5}^{+6.9}$ per year.
This rate is larger than but consistent with that estimated from the PVOL dataset.
The obtained Jovian impact rate is two–three orders of magnitude more frequent than that of terrestrial impacts of the same mass range \citep{Brown2002-aq}.
Also, this is nearly an order of magnitude higher than rates indicated by the surface cratering of Jovian satellites \citep{Zahnle2003-ew,Schenk2004-yg}, 
although there is significant uncertainty due to the small volume of data.

We should note that the total monitoring time of PONCOTS (26.2~h) is much shorter than the estimated impact rate and appears to be inconsistent with each other. As already noted in \citet{Hueso2018-nr}, some individual amateur astronomers observe Jupiter for more than 50 h per yr to find impacts, and only a few of those have observed a flash. 
Furthermore, the DeTeCt program has analyzed observations equivalent to 240 days from more than 150 observers finding only a few smaller impacts than our detected one. 
We conclude that the present detection was made serendipitously during an unlikely (but not impossible) monitoring time. 
Further monitoring of Jovian flashes is required to resolve the inconsistency. 

Jupiter-family comets are thought to be the primary source population of Jovian impactors, and flash detections thus indicate the abundance of $10^6 \-- 10^7$ kg-mass cometary objects and their potential terrestrial flux. According to dynamical models by \citet{Levison2000-ze}, 
the flux of Jupiter-family comets on Earth is $1.0 \times{10}^4$ times lower than that on Jupiter. 
We therefore estimate a terrestrial flux of small-mass comets to be $\sim 1.3_{-1.1}^{+3.1} \times{10}^{-4}$ per year, $\sim 0.2\% \-- 4\%$ of the  flux of near-Earth asteroids of the same mass range \citep{Brown2002-aq, Bland2006-ba}. This range is within the assumed conditions of previous impact flux studies ($<10 \%$, \cite{Bland2006-ba}).
Further monitoring with dedicated telescopes such as PONCOTS will provide more accurate fluxes of Mt-class impacts. 

\subsection{Emission characteristics of megaton-class impacts on planetary atmosphere}

Our detection of an Mt-class impact offers an opportunity for improved understanding of the radiative consequences of Tunguska-class impacts. 
As such impacts occur only once per $10^2 \-- 10^3$ years on Earth (Figure~\ref{fig5}; \cite{Brown2002-aq}), 
their emission characteristics are unknown. Unlike previously observed terrestrial superbolides showing non-thermal line emission-dominant spectra with significant temporal variations \citep{Borovicka1996-op}, 
most observed flash SEDs are consistent with blackbody radiation of uniform temperature (Figure~\ref{fig4}a and Figure~\ref{fig6} in Appendix~\ref{appb}). 
Interestingly, the observed impact flash seems to have a similar spectral trend to that of the Chelyabinsk bolide of 2013, as reported by \citet{Yanagisawa2015-qb}, 
which displayed a blackbody spectrum of uniform temperature during its flare-up phase. 
These trends suggest 
that the impactor suffering ablation is surrounded by a optically-thick shock layer where the bulk of radiation is emitted.
Although shock waves cause the major destruction from outbursts as a result of Tunguska-like impacts \citep{Boslough2008-ae}, the radiation effect also causes considerable burning damage \citep{Jenniskens2019-tg}.
Recent radiation simulations by \citet{Johnston2019-pm} for the Tunguska impact indicate that the radiative flux emitted from the shock layer of typical temperature $\sim 9000$ K would have made a significant contribution to the total radiative energy.
The observed spectral characteristics of the Jovian flash may thus represent radiative features common to superbolides in planetary atmospheres and their consequences. 
Further time-resolved spectral observations of flashes on Jupiter will elucidate the energy-release mechanisms of Mt-class impacts, an understanding of which is necessary for estimation of their potential threat.

\clearpage

\clearpage

\clearpage

\clearpage
\begin{figure}[!pt]
\begin{center}
   \includegraphics[scale=0.85]{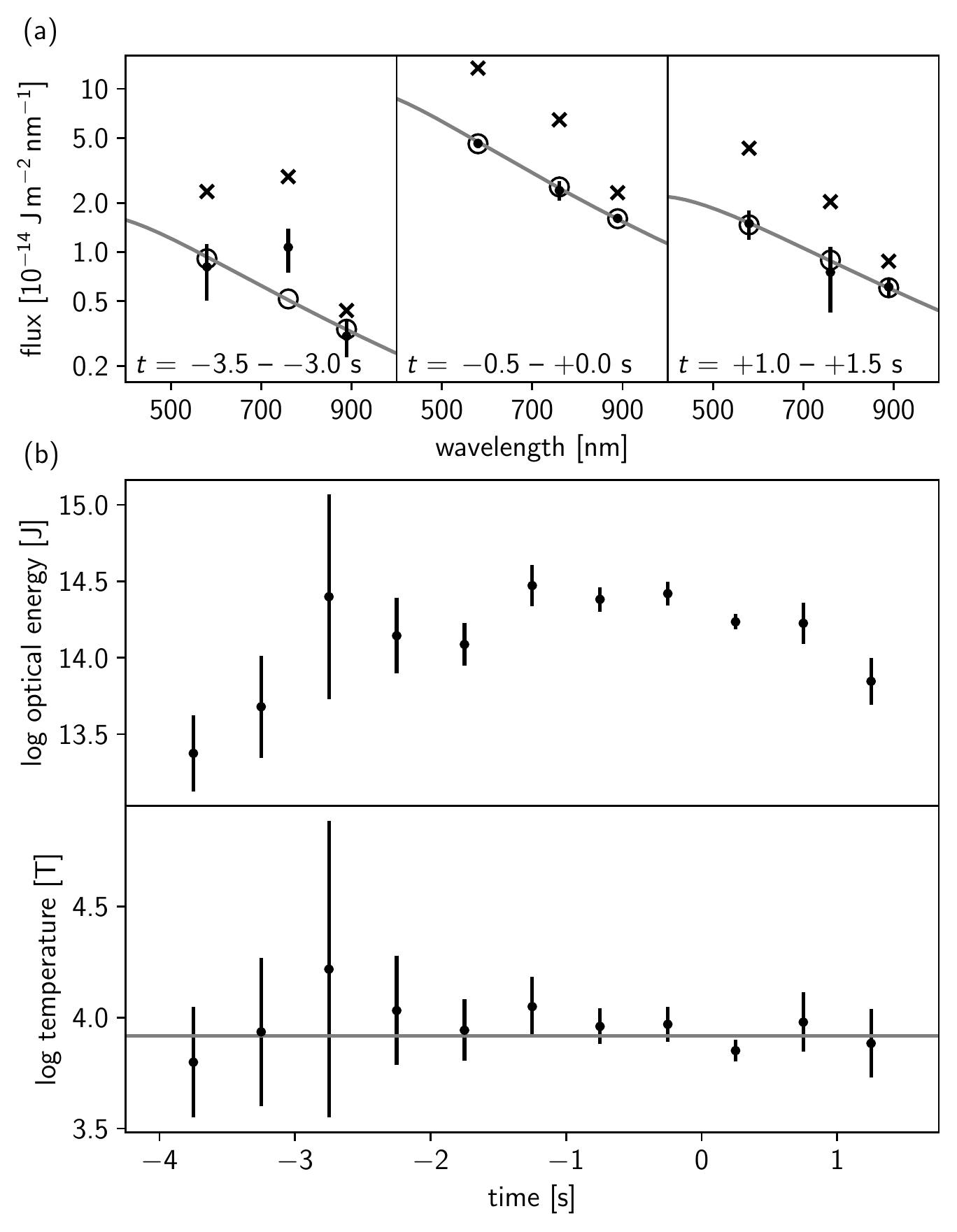}
   \caption{
      Time-resolved spectral energy distributions (SEDs) and model fitting results for the impact flash. 
   (a) Examples of SEDs and best-fit models of 0.5 s bins.
   Left, middle, and right panels show SEDs obtained during the rise, peak, and decay phases, respectively. Crosses and points with error bars represent observed fluxes and those after cloud-reflection correction, respectively. 
   The best-fit spectrum of single-temperature blackbody radiation for each time bin is shown as the grey line.
   Open circles correspond to the model spectrum integrated over the spectral response of PONCOTS photometric bands. 
   (b) Best-fit optical energy (upper panel) and temperature (lower panel) for each 0.5 s bin. 
   Horizontal line in the lower panel represents the weighted-average value of the best-fit temperature (8300 K).
  }
\label{fig4}
\end{center}
\end{figure}

\clearpage
\begin{figure}[!pt]
\begin{center}
   \includegraphics[scale=0.85]{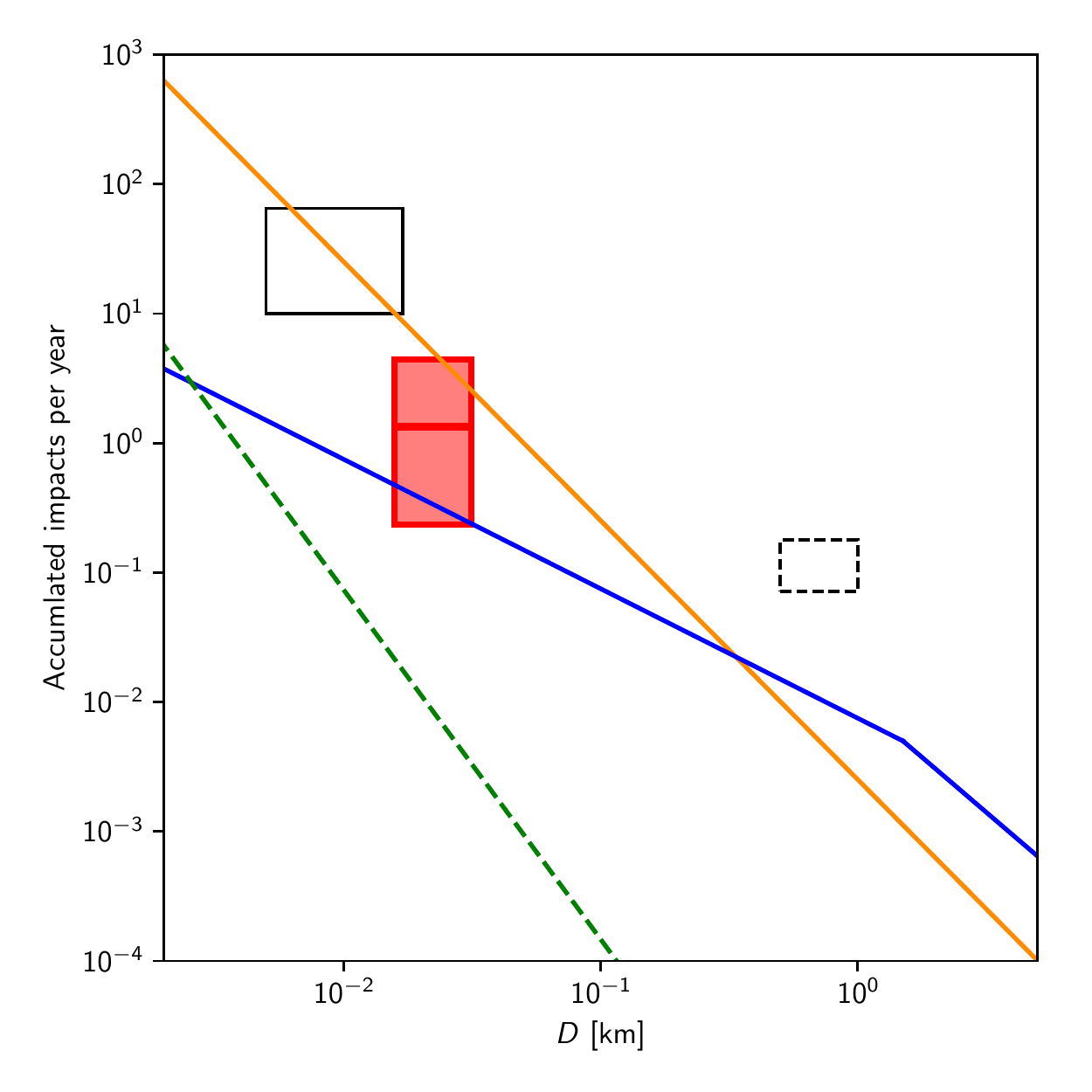}
   \caption{Impact rates on Jupiter and Earth. Our estimated range of Mt-class impact rate is indicated as the red region, with the red horizontal line in this region representing the best-estimate value (1.3 impacts per year). The solid black rectangle indicates the rate based on five detections during $2010\-- 2017$ \citep{Hueso2018-nr}. The dashed black rectangle indicates estimated impact rates on Jupiter based on the Shoemaker–Levy 9 impact of 1994 and impact features observed in 2009 \citep{Sanchez-Lavega2010-rj}. Orange and blue solid lines indicate expected Jovian impact rates from dynamical models of comets \citep{Levison2000-ze} and the cratering record of Galilean moons \citep{Zahnle2003-ew}, respectively. Dashed green line indicates the estimated impact rate on Earth \citep{Brown2002-aq}.
  }
 \label{fig5}
\end{center}
\end{figure}

\clearpage

\acknowledgments
We thank the anonymous referee for a careful review and providing constructive suggestions.
We thank Yasunobu Higa, “{\it yotsu}”, and Victor PS Ang for reporting their observations and providing their data of the impact on Jupiter. We also thank the staff of ALPO Japan for supporting our studies. We thank Erich Karkoschka for providing the data on the methane absorption model, and Liming Li for providing the data on the phase-angle dependence of Jovian albedo. Part of this work was made possible by the development and operation of the JunoCam instrument by Malin Space Science Systems (MSSS).
This research has been partly supported by JSPS grants (18K13606, 21H01153).

\appendix
\section{Details of the flux calibration}
\label{appa}
For the flux calibration, we observed the spectrophotometric standard star HR 7950 on the same night but at a slightly different elevation and time than Jupiter at the impact time. 
We then introduced an additional correction factor to correct the effect of the different conditions.
As the correction factor, we adopt a ratio between the full disk signal values of Jupiter obtained at almost the same time and airmass as the standard star data and those obtained at the impact time. 
We multiply the factor by the observed signal values of the standard star. 
This correction modifies the signal values by less than $10\%$. 
The observed signal values of the impact flash in each frame, $S_{\rm obs}$, were calibrated to derive the calibrated flux, $F_{\rm obs}$, as follows:
\begin{equation}
F_{\rm obs} = \frac{S_{\rm obs}}{S_{\rm SS}} \frac{\int^\infty_0 R(\lambda) \, F_{\rm SS}(\lambda) \, {\rm d}\lambda}{\int^\infty_0 R(\lambda) \, {\rm d}\lambda},
\end{equation}
where $S_{\rm SS}$, $F_{\rm SS}(\lambda)$, and $R(\lambda)$ are the observed signal value of the standard star after the time and airmass correction, the SED of the standard star provided by \citet{Hamuy1992-nw}, and the system response of the PONCOTS band into consideration at wavelength $\lambda$, respectively.

To assess the validity of the present calibration, we compared the calibrated fluxes with those obtained using another method based on a manner of the previous impact flash studies \citep{Hueso2018-nr}.
The calibration was carried out with the full disk brightness of Jupiter by considering the reflected SED of Jupiter and Jupiter-Sun-Earth distances. The Jovian SED is produced by the convolution of the solar spectrum provided by \citet{Colina1996-pv} and reflectively spectrum of Jupiter from \citet{Karkoschka1994-qq}.
The differences between the fluxes calibrated with the standard star and those with Jupiter are $3\%$, $4\%$, and $7\%$ for the V, Gh, and ${\rm CH_4}$ bands, respectively.

\section{SED model fitting results for all time bins}
\label{appb}
\begin{figure}[!pt]
\begin{center}
  \includegraphics[scale=0.85]{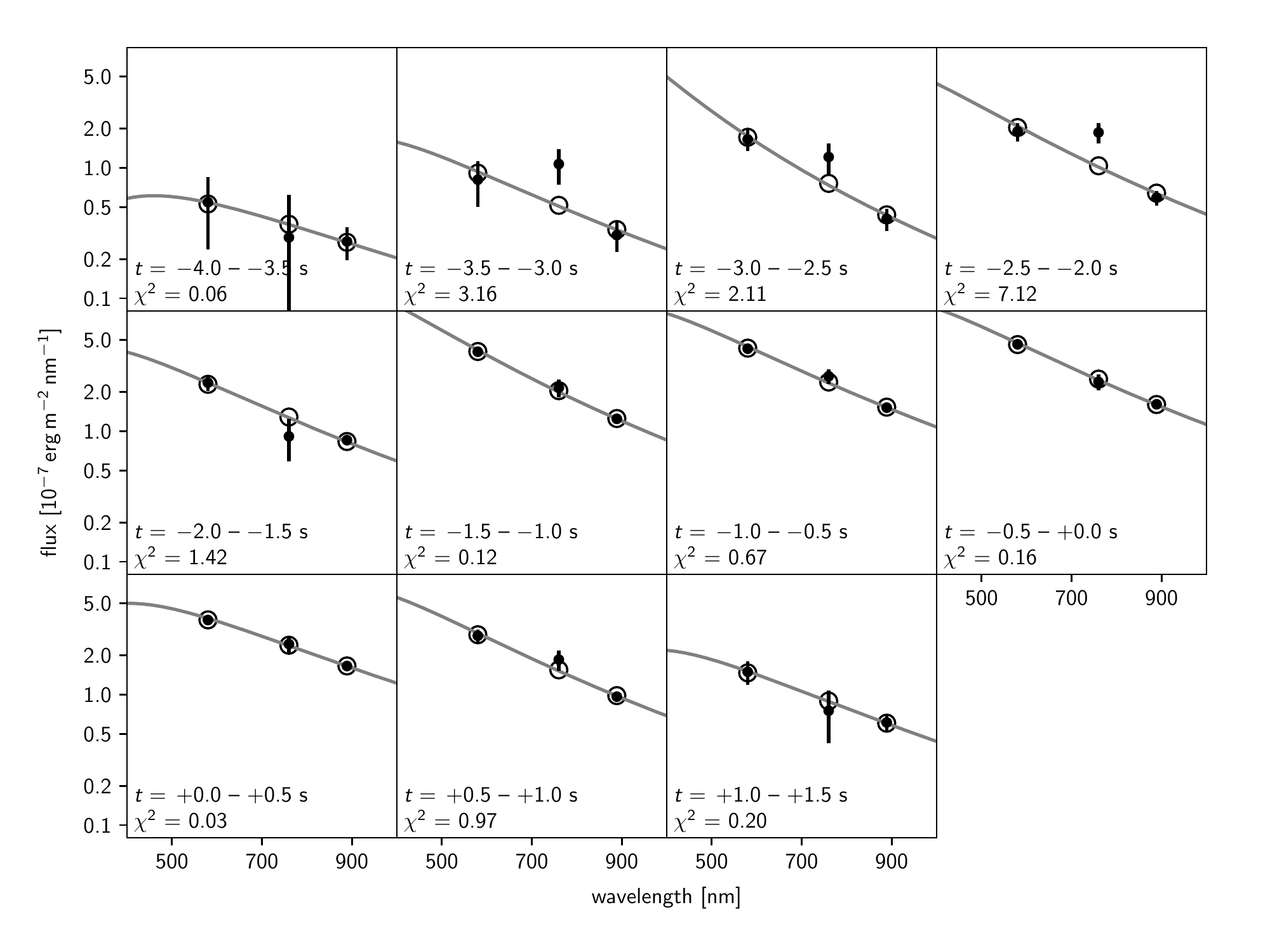}
  \caption{
  Time-resolved SEDs and model fitting results. Each panel shows the observed spectral energy distributions (SEDs) of the impact flash in each 0.5 s bin and the best-fit spectrum of single-temperature blackbody radiation. The points with error bars represent observed fluxes in the three PONCOTS bands after cloud-reflection correction. The best-fit spectrum of single-temperature blackbody radiation for each time bin is shown as a grey line. Open circles represent model spectra integrated over the spectral response of each photometric band. The best-fit $\chi^2$ value from the fit (with one degree of freedom) is also shown in each panel.
  }
 \label{fig6}
\end{center}
\end{figure}

Figure~\ref{fig6} shows the Time-resolved SEDs and model fitting results, as partly shown in Figure~\ref{fig4}a.

\clearpage

\clearpage

\bibliographystyle{aasjournal}
\bibliography{ref01}

\end{document}